# Optical Flow Method for Measuring Deformation of Soil Specimen Subjected to Torsional Shearing

**Piotr E. Srokosz [1], Marcin Bujko [2], Marta Bocheńska [3],* and Rafał Ossowski [4]**

[1]  University of Warmia and Mazury in Olsztyn; psrok@uwm.edu.pl
[2]  University of Warmia and Mazury in Olsztyn; marcin.bujko@uwm.edu.pl
[3]  University of Warmia and Mazury in Olsztyn; marta.baginska@uwm.edu.pl
[4]  Gdansk University of Technology, rafal.ossowski@pg.edu.pl
*   Corresponding author: marta.baginska@uwm.edu.pl

**Abstract:** In this study optical flow method was used for soil small deformation measurement in laboratory tests. The main objective was to observe how the deformation distributes along the whole height of cylindrical soil specimen subjected to torsional shearing (TS test). The experiments were conducted on dry non-cohesive soil specimens under two values of isotropic pressure. Specimens were loaded with low-amplitude cyclic torque to analyze the deformation within the small strain range (0.001–0.01%). Optical flow method variant by Ce Liu (2009) was used for motion estimation from series of images. This algorithm uses scale-invariant feature transform (SIFT) for image feature extraction and coarse-to-fine matching scheme for faster calculations. The results were validated with the Particle Image Velocimetry (PIV). The results show that the displacement distribution deviates from commonly assumed linearity. Moreover, the observed deformation mechanisms analysis suggest that the shear modulus *G* commonly determined through TS tests can be considerably overestimated.

**Keywords:** granular materials; ; small displacement measurement; torsional shear; SIFT optical flow; particle image velocimetry.

## 1. Introduction

Soil as a natural granular material displays complex mechanical behavior. It can react to a loading differently depending on many factors like grains sizes and shapes, density or moisture content. Despite soil's granular composition it is not uncommon that for structural design purposes it is considered to behave like a continuous material [2-4]. For simplification many models assume that some initial part of soil deformation path is purely elastic [5-9]. As most of the soil-structure interactions are within the small strain range (0.001 – 0.01%) [10,11] this simplification is very useful and it can even be a satisfactory approximation for simple and typical engineering problems. However, research shows that even the initial deformation part in soil can be more complex than that [12-15] and the soil response to stress in small strain range is not yet properly described.

To learn more about the nature of the soil mechanical response research needs to be oriented on very precise laboratory testing and observation of processes taking place in specimen during the test. Yet devices for soil mechanical laboratory testing usually allow to measure only the resultant value of a physical quantity e.g. specimen deformation. And those resultant values are often a basis for further soil parameter evaluation (like soil stiffness parameters). As far as that approach is easy to use and accessible even for non-experts, the tests do not provide us with any information about material mechanical behavior inside the specimen. It is a known fact that different stress distributions can give the same resultant value measured near the boundary [16] and deformation at one point of the material specimen does not indicate the deformation distribution within the whole sample. To take a measurement inside a soil specimen sensors can be attached to the specimen locally inside the laboratory testing device. However, in order to learn more about the distribution of measured value one should rather use methods like X-ray computed tomography, magnetic resonance imaging (which in most cases is not economically justified) or various optical methods [17-20].

Due to the extreme importance of observation of deformations of soils tested in geotechnical laboratories, research in this field is carried out using contact and non-contact measurement techniques. Among them there are techniques for investigating soil displacement patterns [21], wireless chemiluminescence-based sensors [22] fiber Bragg grating sensors [23], particle image velocimetry (PIV) and photogrammetry [24].

In this study an optical flow method variant is used. The optical flow is a visual experience caused by the relative motion of the observer and objects in the environment. The phenomenon is most often defined as "the distribution of apparent velocities of movement of brightness pattern in an image" [25]. The optical flow concept came from the studies on animal's visual perception by American psychologist James J. Gibson in the 1950s [26]. The apparent motion of objects in the animal's field of view provides information about the spatial arrangement of the objects in space and the rate of change of this arrangement [25].

The research on the optical flow applications focuses strongly on motion estimation and video compression aspects. It is widely used in robotics for object detection and tracking or movement detection [27]. As the optical flow concept is a major aspect of a human (animal) vision, the technique is also adopted for development of machine vision [28]. The optical flow-based techniques are intensively developed and used for a wide spectrum of different purposes like fault detection in production processes [29], intelligent vehicle navigation [30], micro air vehicles control [31], structural displacement monitoring [32] and many more (see [33-35]).

A validation of the final optical flow results is performed using an alternative image processing method, Particle Image Velocimetry (PIV). PIV is commonly used for measuring velocities in fluids but it also has been proven effective for a variety of soil mechanics problems [36-38]. The PIV method was probably first mentioned by Adrian in 1984r. [39]. In the last four decades both the imaging techniques and the calculation algorithms were being intensely modified [40-45].

Soil specimens' deformations during the Torsional Shearing (TS) test are examined. Torsional shearing of soil cylindrical specimens is a well-known soil testing method commonly used in geotechnical engineering from the 1930s. Initially tests based on torsional shearing were performed mainly to obtain information about mechanical properties of soils and rocks subjected to cyclic loads, especially seismic loads. The resonant frequency of the specimen subjected to torsional vibrations allows for estimation of shear modulus $G$ and damping ratio $D$. Those parameters can be evaluated using the elastic wave propagation theory. Laboratory devices for performing such tests are called resonant columns (RC) and were invented in Japan by Ishimoto and Iida [46] and (independently) in the USA by Birch and Bancroft [47]. Since the 1960s the resonant column has been widely used in studies of dynamic behavior of soils and rocks. In the 1990s, modifications in the device control system led to the invention of torsional shear (TS) apparatus. The TS device has found wide application in soil research under low frequency cyclic load conditions [48]. Currently, geotechnical laboratories around the world are equipped with a combined version of the RC and TS devices called multifunction RC/TS apparatus. Such a device is widely used for dynamic characterization of soils, especially for damping and stiffness measurements [49-53], applying nonlinear vibration analysis [54], the study of special granular materials (as micaceous sands [55]) and studies including hardware modifications (i.e. for the large strains [56] or permanent deformations in small strain range [57]). The originally modified construction of RC/TS apparatus working in TS mode is used in the research presented in this paper.

This paper present the results of soil displacements measurement with SIFT optical flow method. The PIV method is used for validation. Dry non-cohesive soil cylindrical specimens were subjected to low-amplitude cyclic torque, so that small strains could be analyzed. The main focus of the conducted experiments was to observe the displacements distribution along the whole specimen's height and verify the linearity assumption.

## 2. Materials and Methods

### *2.1. Experimental setup*

The experimental setup consisted of a torsion shear apparatus WF8500, a 5 megapixel digital camera installed on the ARAMIS 5M (system for high-precision motion measuring) and 1000 W halogen lamp (Figures 1 and 2).

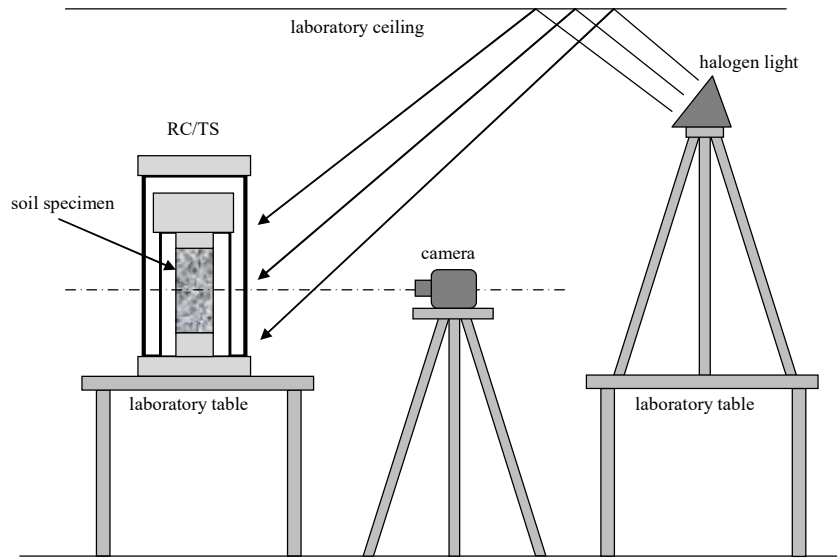

**Figure 1.** The experimental setup scheme

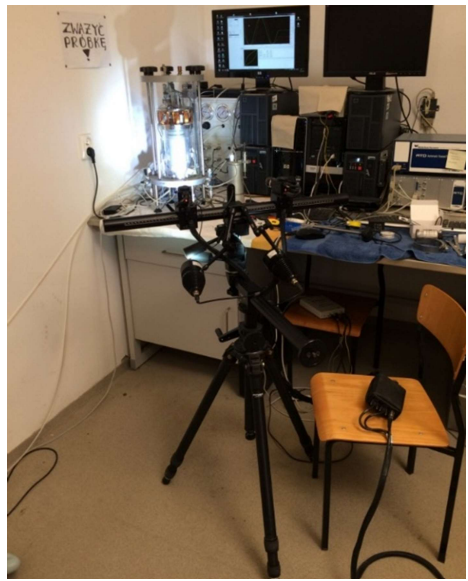

**Figure 2.** The specimen in TS apparatus directly illuminated with LED lamp to focus the camera.

Photos were taken at one second intervals. Despite all efforts, the digital image correlation built into the ARAMIS system was unusable due to interference created by two cylinders made of polycarbonate that isolated the soil specimen from the environment. Another important issue was the optimal illumination of the lateral surface of the specimen. The lighting had to be diffused so that it would not cause reflections from the walls of the polycarbonate cylinders. It was found that the best way to ensure optimal conditions for taking pictures is to illuminate the specimen with light reflected from the white ceiling of the laboratory. The light can be diffused by other methods (like on-camera flash diffusers, softboxes, photography umbrellas) to ensure uniform non-direct illumination of the specimen. However, to focus the camera, the specimen was temporarily illuminated directly by the LED lamp (Figure 2).

*2. 2. Torsional Shearing device*

The RC/TS (Resonant Column/Torsional Shearing) device used in this study is produced by Wykeham Farrance (model WF8500). Figure 3 presents the components of the device.

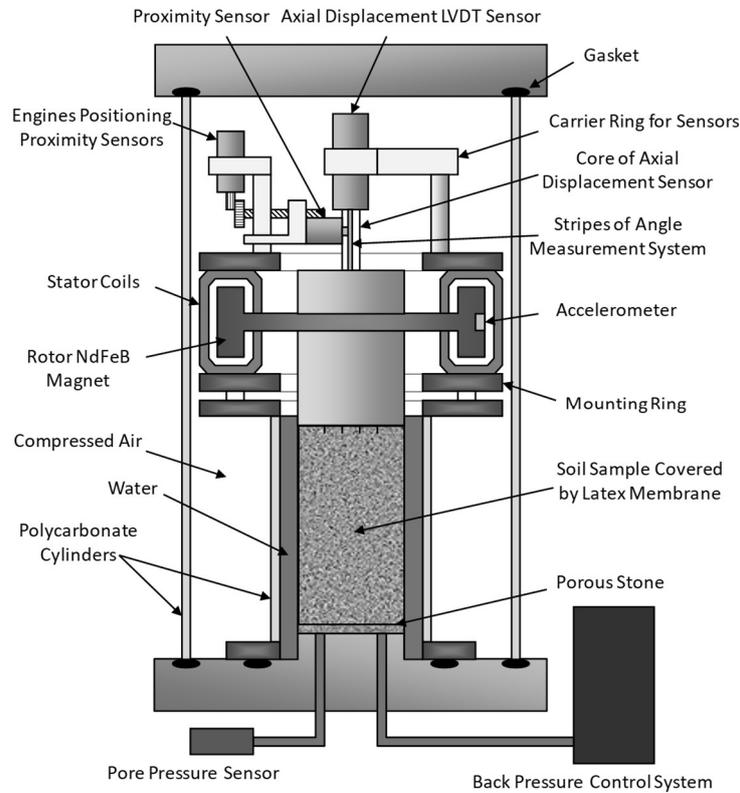

**Figure 3.** Components of RC/TS apparatus [58].

In TS mode a cylindrical soil specimen is subjected to torque *T* which changes harmonically. It results in a specimen twist measured by proximity transducers (Figure 4). The load frequency is below 1 Hz (typically 0.01 Hz, in modified version of the motor driver/controller even $10^{-5}$ Hz including different torque excitations signals like sinusoidal, trapezoidal, triangular-saw etc.).

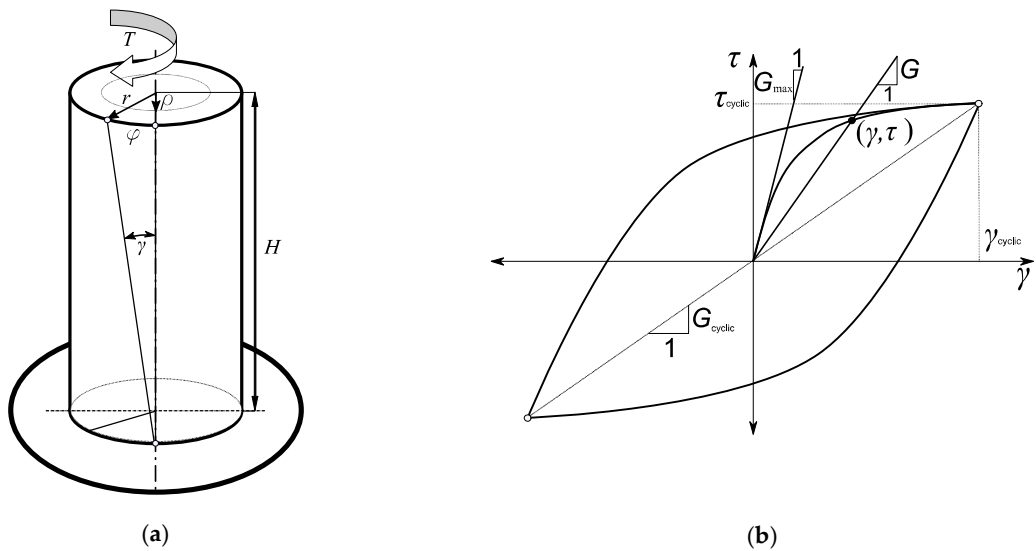

(a) (b)

**Figure 4.** (**a**) Torsional shearing of cylindrical soil specimen (*r* - specimen's radius, $\rho$ - specimen's reduced radius, *H* - specimen's height, $\varphi$ - twist angle, $\gamma$ - shear strain, *T* - torque); (**b**) Theoretical hysteresis loop of shear stress-strain relationship in cyclic shearing test ($\tau$ - shear stress, *G* - Kirchhoff's modulus) [59]. $G_{max}$ - initial shear modulus, $G_{cyclic}$ - shear modulus obtained from cyclic TS test, $\tau_{cyclic}$ - maximum shear stress obtained from cyclic TS test, $\gamma_{cyclic}$ – maximum shear stress obtained from cyclic TS test.

The measured torque and twist angle allow calculation of tangential stress and shear strain which are then used for Kirchhoff's modulus *G* estimation. Precise control of small torque amplitudes allows for soil behavior examination in small strain range (0.001 – 0.01%) which is typical for commonly encountered soil-structure interactions [10,11].

The reduced radius ϱ of the specimen (Figure 4(a)) is defined as follows

$$\rho = \kappa r, \quad (1)$$

where
*r* is the radius of the specimen,
$\kappa$ is a coefficient dependent on shear strain $\gamma$ range; $\kappa$=0.8 for $\gamma$<0.001%, $\kappa$=0,65 for $\gamma$≈0.1% according to ASTM D4015-92 standard [60]; $\kappa$=0.67 is a value recommended in the RC/TS apparatus manual and used in our study.

In the torsional shearing mode, the cylindrical specimen is cyclically loaded with torque *T* of a harmonically changing amplitude $T_0$. The material's response, i.e. the twist angle $\varphi$, is measured with the proximity sensors. The cyclic torque frequency *f* is set lower than 0.1 Hz, so the torque should not be considered as a dynamic load, but as a slow-changing cyclic load. Basing on the parameters controlling the torque value and the corresponding value of the specimen twist angle, the device software determines the tangential component $\tau$ of the stress state and the shear strain $\gamma$. The shear modulus G is calculated using the following equation

$$G = \frac{\tau(\gamma_{max})}{\gamma_{max}}, \quad (2)$$

where

$$\gamma(\varphi) = \frac{\rho \cdot \varphi}{H}, \quad (3)$$

and

$$\tau(T) = \frac{\rho \cdot T}{I}, \quad (4)$$

with

$$T(t) = T_0 sin(\omega \cdot t), \quad (5)$$

*H* – specimen's height, $\rho$ – specimen's reduced radius, *I* – second area moment of the specimen's horizontal cross section, $\omega$ - angular frequency.

Figure 4(b) shows an idealized stress–strain relation obtained for cyclic load (symmetric periodic waves) of a predetermined amplitude. The obtained stress–strain relations form a hysteresis loop. The soil's mechanical response in form of the hysteresis loop indicates that the material accumulates elastic energy of reversible deformations but loses some energy due to damping. The lost part of energy is equal to the work done by applied external force (the torque). What's important is that the results interpretation is based on the assumption that the soil reacts like a continuous viscoelastic material. During the test the shearing angle is measured only in the specimen's cross-section near its upper end. The deformation is assumed to be linearly distributed along the specimen's height. However, there is no evidence that the specimen always deforms proportionally over its entire height.

Figure 5(b) presents a soil specimen (from previous study) subjected to torsional shear (maximum rotation stage). There are 7 phosphorescent markers placed near the specimen's surface along the specimens' height. Visible marker's displacements (circumferential displacements due to applied torque) indicate that displacement distribution along the height of the specimen may not be linear. Moreover, the markers placed near the lower end (fixed) seem not to undergo any displacement. In this study a method of surface displacements measurement using SIFT optical flow is proposed to verify this preliminary observation

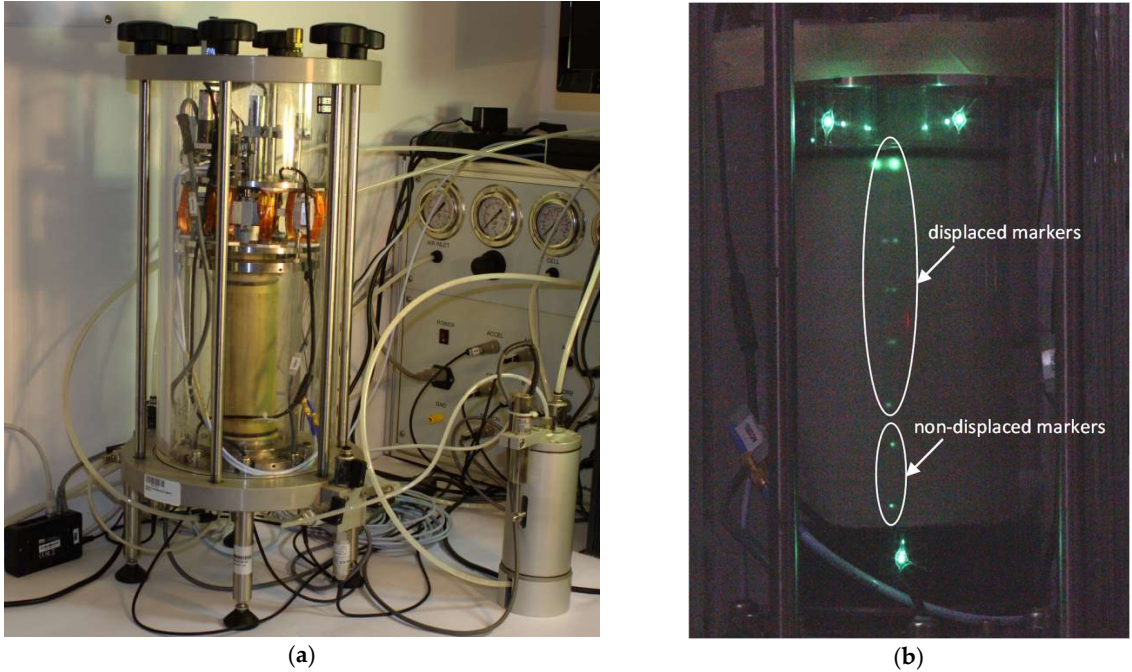

(a)        (b)

**Figure 5.** (**a**) The TS device; (**b**) Result of optical observation of phosphorescent markers' displacements during TS test with visible non-displaced points [61].

## *2. 3. Materials used in tests*

A few types of granular, non-cohesive materials were selected for the study, which were prepared according to [62] (the specimen was prepared from pre-formed non-cohesive soil fractions). All specimens were made of soils that according to ISO 14688-1:2002 standard [63] can be classified as sands (Sa). The soils used for specimen preparation are different mixtures of dry coarse (CSa, 0.63-2.0 mm), medium (MSa 0.2-0.63 mm) and fine (FSa, 0.063-0.2 mm) grained silica sand. Seventeen mixtures with different parameters and grain size distributions were tested for each set of testing conditions. According to [64] selected soils have been classified as suitable for testing in Torsional Shearing apparatus. For a greater clarity the results for specimens made of one type of soil are presented (results for the other types do not differ much in terms of quality). The properties of soil selected for further analyses are summarized in Table 1 : the values of the particle diameter at 50%, 60%, 10% and 90% in the cumulative distribution ($d_{50}$, $d_{60}$, $d_{10}$, $d_{90}$), void ratio (*e*), uniformity coefficient ($C_U$), soil density ($\rho$), specific gravity ($G_s$). The shape of grains is shown in Figure 6. Figure 7 presents grain size distribution of tested material.

**Table 1.** Parameters of tested soil

| Soil type | $G_s$ [-] | $d_{50}$ [mm] | $d_{60}$ [mm] | $d_{10}$ [mm] | $d_{90}$ [mm] | $C_U$ [-] | *e* [-] | $\rho$ [g/cm³] |
|---|---|---|---|---|---|---|---|---|
| silica sand | 2.65 | 0.33 | 0.35 | 0.22 | 0.55 | 1.6 | 0.50 | 1.76 |

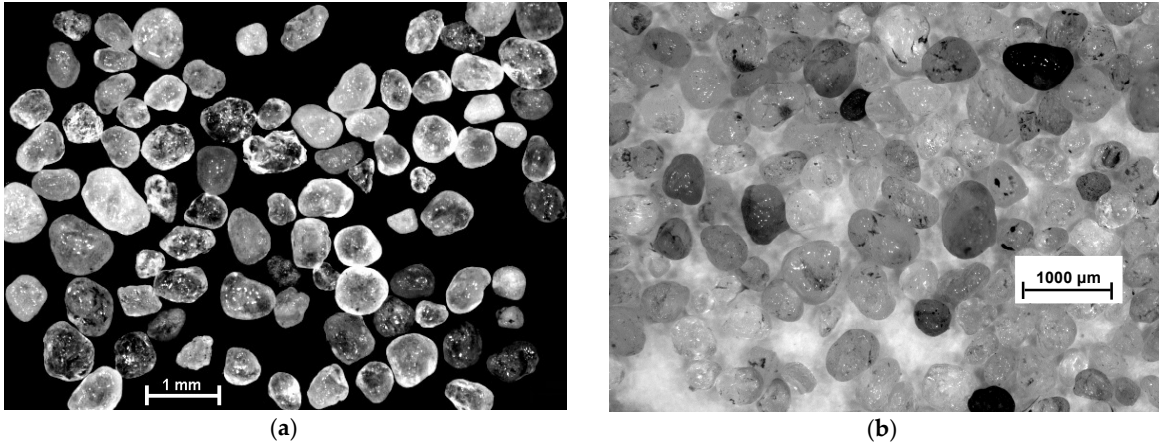

(**a**)　　　　　　　　　　　　　　　　　　(**b**)

**Figure 6.** Tested material. Grains in magnification (**a**) on a black background; (**b**) on a white background.

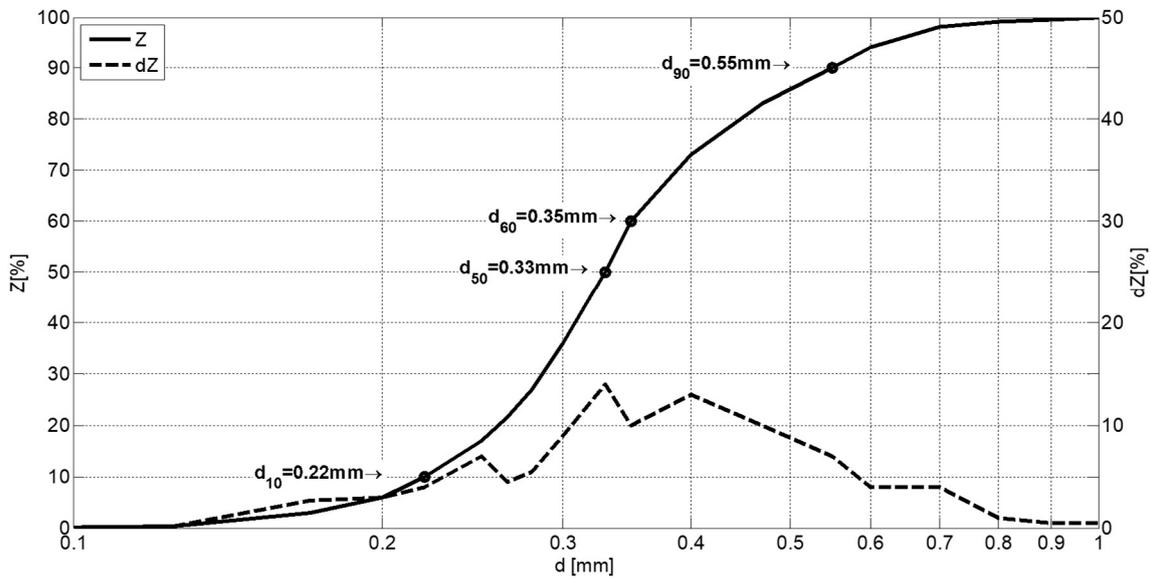

**Figure 7.** Grain size distribution in the tested material (*d* - equivalent particle diameter, *Z* - cumulative less than indicated size, *dZ* - volume share per particle size interval).

## *2. 4. Optical flow estimation*

In this section we describe in detail the brightness optical flow [65] and the Scale Invariant Feature Transform (SIFT) used in this study. The brightness optical flow is a technique for determining the displacement field using recognition of spatiotemporal patterns of image intensity. The SIFT algorithm used for local feature detection allowed to track very small displacements with a good accuracy.

### *2. 4.1. Brightness optical flow*

The optical flow methods use time-ordered image sequences to estimate the motion (velocities or discrete displacements) of objects in the image series. The methods use partial derivatives with respect to the spatial and temporal coordinates to calculate objects' current velocity.

To describe the motion between two image frames (taken at points t and $t + \Delta t$) in time at every position on a 3D grid the optical flow methods use Taylor series approximations of the object motion trajectory.

In this study the brightness optical flow was used for 2D image analysis captured with a single digital camera. The points located on cylindrical side surface of the specimen were analyzed. The displacement fields were obtained by projecting changes in the points' positions in 3D to 2D (considering the specimen's geometry, camera position and light refraction caused by external polycarbonate cylinders). Although it was not the focus of this study, the method allows to recreate the 3D displacement field basing on a sequence of 2D images (see [66] for detailed procedure).

In camera-centered coordinates each point in a 3D space moves along a 3D path. When projected onto the image plane each point moves along a 2D path. Considering the 2D+$t$ case, where image pixels with coordinates $(x_1, x_2, t)$ and intensity $I(x_1, x_2, t)$ move by $\Delta x_1, \Delta x_2$ over time $\Delta t$, a new image is obtained $I(x_1 + \Delta x_1, x_2 + \Delta x_2, t + \Delta t)$. It is assumed that pixel intensities are constant between the frames (the brightness constancy constraint)

$$I(x_1, x_2, t) = I(x_1 + \Delta x_1, x_2 + \Delta x_2, t + \Delta t), \tag{6}$$

as the movement is assumed to be small. Removing higher order terms from the Taylor series

$$\begin{aligned}I(x_1 + \Delta x_1, x_2 + \Delta x_2, t + \Delta t) \\ = I(x_1, x_2, t) + \frac{\partial I}{\partial x_1}\Delta x_1 + \frac{\partial I}{\partial x_2}\Delta x_2 + \frac{\partial I}{\partial t}\Delta t + higher\ order\ terms\end{aligned} \tag{7}$$

we get

$$\frac{\partial I}{\partial x_1}\Delta x_1 + \frac{\partial I}{\partial x_2}\Delta x_2 + \frac{\partial I}{\partial t}\Delta t = 0 \tag{8}$$

Dividing (7) by $\Delta t$ we obtain the optical flow equation

$$\frac{\partial I}{\partial x_1}u_1 + \frac{\partial I}{\partial x_2}u_2 + \frac{\partial I}{\partial t} = 0, \tag{9}$$

where $u_1 = \Delta x_1/\Delta t$ and $u_2 = \Delta x_2/\Delta t$ are the components of the velocity, $\partial I/\partial x_1, \partial I/\partial x_2$ and $\partial I/\partial t$ are the image gradient components along the horizontal axis, the vertical axis, and time.

As there are two unknowns $(u_1, u_2)$ and only one equation (9) the problem cannot be solved without introducing additional constraints. To address this issue the optical flow methods are using different additional equations (e.g. global smoothness constraint, known as Horn-Schunck method [25] or constant flow in a local neighborhood, known as Lucas-Kanade method [67]).

Apart from the Gradient-Based methods many variants of optical flow estimation are used to approximate motion field from time-varying image intensity including correlation, block-matching, feature tracking or energy based methods. Despite the differences between optical flow techniques, many of them generally follow three processing stages listed below [68].
1. Pre-filtering or smoothing with low-pass/band-pass filters in order to extract signal structure of interest and to enhance the signal-to-noise ratio.
2. Extraction of basic measurements such as spatiotemporal derivatives (to measure normal components of velocity) or local correlation surfaces.
3. Integration of these measurements to produce a 2D flow field which often involves assumptions about the smoothness of the underlying flow field.

A major drawback of the Gradient-Based Estimation method is its sensitivity to conditions commonly encountered in real imagery. Highly textured regions, motion boundaries and depth discontinuities might be troublesome. In practice, for the observed brightness (image intensity) of any object point to be constant over time the following conditions need to be satisfied
- surface radiance remains fixed from one frame to the next,
- non distant point source (object),
- light distance is constant or changing the distance has no effect,
- no object rotations,
- no secondary illumination.

In experiments performed in this study the aforementioned conditions were satisfied.

*2. 4.2. Scale-Invariant Feature Transform method (SIFT)*

Scale Invariant Feature Transform is an algorithm for local feature detection in images. The method was originally developed by David G. Lowe in 1999. It is used in computer vision for image matching and object recognition. The information about the local image features extracted by SIFT algorithm is invariant to scale, translation and rotation in the image domain [69]. SIFT can be used to measure velocities, displacements and deformations based on the analysis of local features of recorded image series. The method consists in finding similarities in the analyzed images basing on distinguished features. Those features are characteristic patterns of pixels, usually found on high contrast regions like edges and peaks.

The SIFT algorithm starts with selection of potential location for finding features. When the feature keypoints are localized the rotation invariance is achieved by assigning orientation to the keypoints. The obtained keypoints are recorded as a high dimensional feature vectors called SIFT keys. This approach transforms an image into a large collection of the SIFT keys. Each of them is invariant to image translation, scaling, 3D projection. As author highlights [69] method is also less sensitive to projective distortion and illumination change, which is a significant advantage over other similar methods.

Ce Liu in his Ph.D. thesis in 2009 [1] proposed a SIFT flow method which was formulated just like the optical flow but instead of the pixels' RGB values the SIFT descriptors were used for matching between image frames. SIFT method was used for feature extraction purposes. Every pixel was assigned the 128-D vector containing information about the pixel's nearest neighbors.

When matching two "SIFT images" (features) $s_1$ and $s_2$ the goal is to minimize an energy function $E$ of displacement $\boldsymbol{w}(\boldsymbol{p}) = (u(\boldsymbol{p}), v(\boldsymbol{p}))$, where $\boldsymbol{p} = (x, y)$ is a pixel. The $u(\boldsymbol{p})$ and $v(\boldsymbol{p})$ values are limited to integers so the number of possible states is significantly reduced. The following form of the energy function is introduced (similar to optical flow)

$$E(w) = \sum_{\boldsymbol{p}} min\left(\left\|s_1(\boldsymbol{p}) - s_2(\boldsymbol{p} + \boldsymbol{w}(\boldsymbol{p}))\right\|_1, d_1\right) + \sum_{\boldsymbol{p}} \eta(|u(\boldsymbol{p})| + |v(\boldsymbol{p})|) \\ + \sum_{(\boldsymbol{p},\boldsymbol{q}) \in \varepsilon} min(\alpha|u(\boldsymbol{p}) + u(\boldsymbol{q})|, d_2) + min(\alpha|v(\boldsymbol{p}) + v(\boldsymbol{q})|, d_2) \tag{10}$$

where $\varepsilon$ is a set of neighborhoods of $\boldsymbol{p}$ and $d_1$, $d_2$ are thresholds. The energy function consist of three components. The first term is for finding pixels with similar SIFT descriptors in the next frame. The second term ensures that pixels found are close to $\boldsymbol{p}$ (i.e. $\boldsymbol{w}(\boldsymbol{p})$ is as small as possible). And the third term constraints the set of solutions so that flow of pixels from neighborhood of $\boldsymbol{p}$ is similar to $\boldsymbol{w}(\boldsymbol{p})$ (smoothness term).

The method consists in the successive search for all possible variants of displacement of the analyzed pixel structure which can take a lot of time (more than 2h for a pair of 256×256 images with a 16GB memory usage [1]). Ce Liu has simplified the algorithm by applying coarse-to-fine matching scheme. Instead of analyzing all the possibilities of movement in detail, it is searches for larger structures (coarse) and for similarities (fewer opportunities) first, and then for the details (fine). This approach significantly shorten the time of the necessary calculations (for the same pair of images it took about 30 s on a two quad-core 2.67 GHz CPUs and 32 GB memory [1]).

The coarse-to-fine matching not only speeds up the process but also minimize energy function better in most cases compared to the ordinary scheme [1].

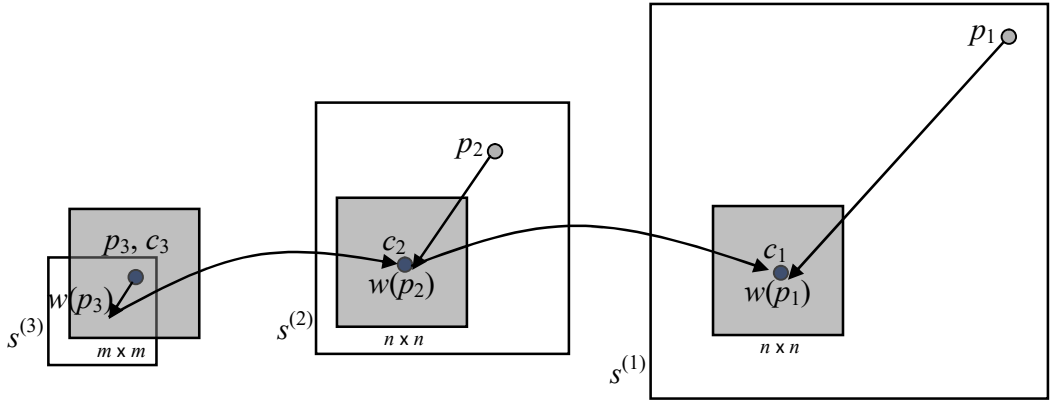

**Figure 8.** The coarse-to-fine SIFT flow matching scheme [1].

The idea of coarse-to-fine SIFT flow matching is illustrated in Figure 8. The searching for the best match starts at the least detailed image version $s^{(3)}$. Lowered image resolution makes the general direction of movement easier to find. Also the search window $c_3$ (the gray square) centered on the point $p_3$ is relatively large compared to the image size $m \times m$. The flow vector $w(p_3)$ is propagated to more detailed images $s^{(2)}$ then $s^{(1)}$ and it can be more precisely determined with every step. Searching the whole area of the original picture $s^{(1)}$ would be far more time-consuming. This procedure significantly lowers the level of computational complexity what results in faster calculations.

The source code of Ce Liu's SIFT flow is publicly available [1]. Since the program needs some human assistance with managing the image layers and objects properly, the graphical user interface allows the user to easily specify some of the key aspects of motion analysis. In more detail, the user can define layer configurations, objects contours, specify the depth of each object. The user needs to specify those at one or several key frames and the program automatically tracks the contours for the rest of the frames.

Author mentions that object tracking process can go wrong unless the occlusion is modeled properly. The system allows the user to change the depth of the object which is automatically interpolated with a smooth depth function. HSV color space (see Figure 9 and interesting HSV manipulations in [70]) has been applied to indicate the depth, by fixing S (saturation) and V (value) as 255, and letting H (hue) reflect the depth value. Warmer color (closer to red) indicates smaller value of depth, closer to the camera, whereas colder color (closer to blue) indicates larger value of depth, further of the camera.

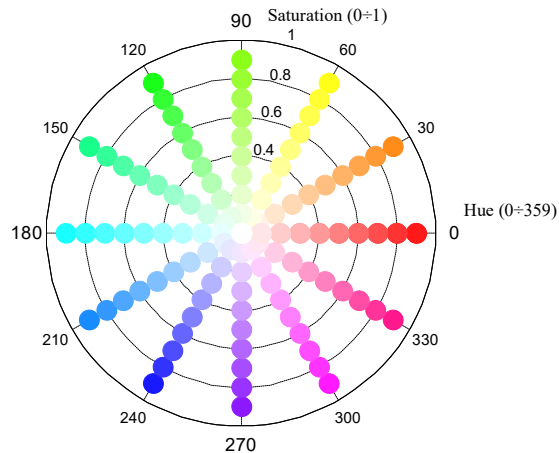

**Figure 9.** HSV color space used for occlusion handling (for value V=1).

*2.5. Particle Image Velocimetry (PIV)*

In this study a PIV numerical implementation by Thieckle and Stamhuis [71] known as DPIV (Digital PIV) was used for results validation. The main part of DPIV analysis is the determination of a correlation matrix *C*. The matrix is obtained from comparing a pair of images, *A* and *B*. The images comparison is done using interrogation areas (IA) of chosen dimensions. Discrete cross correlation (DCC) algorithm compares the IA pairs and the result is a matrix of the most probable displacements (number displaced points=number of IAs)

$$C_{mn} = \sum_{i,j} A_{ij} B_{i-m, j-n}. \tag{11}$$

The equation can be solved with direct cross correlation [72], particle image pattern matching [73], or convolution filtering [74]. The SIFT optical flow results validation was performed using one of the most effective PIV algorithms based on discrete Fourier transform (DFT) [71,75]. In the analysis the following algorithm configuration variants were considered

- 2 passes with IA dimensions 64 and 32 pixels (experiment marked as 2p64)
- 2 passes with IA dimensions 8 and 4 pixels (marked as 2p8)
- 4 passes with IA dimensions 32 pixels, refined to 16, 8 and 4 pixels (marked as 4p32)

*2.6. Specimen preparation*

A cylindrical soil specimen was formed using a template with latex membrane installed inside it. Each cylindrical specimen was compacted with a hand rammer in five layers of dry sand, taking care not to damage the texture covering the lateral outer surface of the membrane (Figure 10). Each specimen had diameter $D = 70\ mm$ and height $H = 140\ mm$.

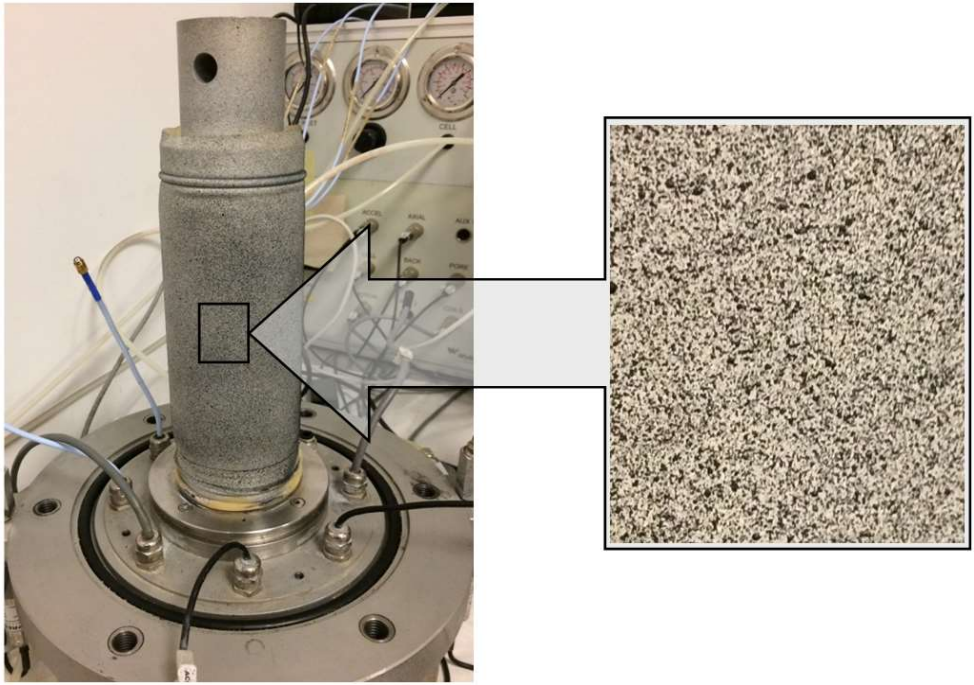

**Figure 10.** Soil specimen in latex membrane covered with fine pattern

Optical displacements tracking requires a pattern consisting of light and dark points on the specimen's surface (see Figure 10). To satisfy this requirement the texture covering the membrane was prepared by first applying a thin primer of white paint and then gently spraying black paint, trying to get the highest contrast between the neighboring points. It should be noted that the paint layers were so thin that they did not change the mechanical properties of the latex membrane. However, it was noted that despite the use of waterborne paints, long-term contact of the paint (over 2 months) led to chemical degradation of latex. Therefore, the membranes were prepared immediately before each test.

The experiment steps and the duration of each step were as follows
1. Shaping the soil specimen in latex membrane and cylindrical form (15 min)
2. Spraying the quick drying paint over the membrane. First layer, a white base, and second layer, a black texture (2-3 min with 20 min break for drying)
3. Assembling the RC/TS device chamber equipment, i.e. internal cylinder, drive system, measurement system, external cylinder (20-30 min)
4. Pressurizing the chamber (1 h)
5. Loading the specimen with torque and measuring the displacements (5 h)
6. Repeating the test for different torque amplitudes to ensure the effecting displacement is within the small strain range (2-4 iterations with a night break)

The total duration of a single experiment did not exceeded 100 h. As it was noted, after 2 months the painted latex membranes were disintegrating into pieces. To ensure the paint did not affect mechanical properties of latex membrane during the experiment time the relaxation test was conducted on two latex stripes of width 12 mm (one cover with paint and one with original clean surface) stretched with a constant force approx. 7.5 N. After 110 h similar force drop was observed in both cases. The degradation of latex mechanical properties during the time that the main experiments took was not noticeable.

Weight and geometrical dimensions of formed specimens were precisely controlled and meticulously recorded. Due to the large number of specimens prepared this way in the last few years, the reproducible mechanical features of all the prepared specimens were obtained.

After the specimen was formed inside the TS apparatus, the soil was subjected to an isotropic pressure (0.2-30 kPa) for one hour. Then, a cyclic twisting test was carried out with the test parameters shown in Table 2. Seventeen tests were performed for each set of testing conditions. However, two examples of tests were selected to present in detail in this paper. The first test was carried out at isotropic pressure close to zero (i.e. close to atmospheric pressure) to ensure noticeable residual deformation (specimen label "A"). The second test was performed at pressure corresponding to the natural conditions of the bed from which sand was collected (specimen label "B"). Tests were carried out with the minimum frequency to ensure that load (torque) can be considered as nearly static (slow-changing) and not as dynamic. The minimum frequency that can be set with TS device software is 0.01 Hz.

In our study we focused on first cycle to observe the reaction of soil that was not previously loaded (no loading history). SIFT optical flow was used to detect possible local deformation anomalies during the initial phase of loading. Thus the results for maximum and minimum torque amplitude are obtained during the first cycle. The results for the residual state were obtained after 3 cycles and the additional observation time (permanent soil specimen' deformation). The observation time was 4.4 h so the whole experiment was concluded in 5 h period.

All data for the optical flow information could have be obtained from only one cycle of test (plus the observation time) however, the standard software of the WF8500 apparatus makes it impossible to conduct a test with fewer than 3 cycles. Thus, the minimum number of cycles was set. The load amplitude was selected by trial and error method (2-4 iterations) so that the displacements were in the desired strain range (0.001-0.01%) for each specimen.

Table 2. TS tests parameters

| Specimen label | Confining pressure [kPa] | Amplitude [V] | Frequency [Hz] | Number of cycles [-] | Observation time after the test [h] |
|---|---|---|---|---|---|
| A | 0.2 | 1.0 | 0.01 | 3 | 4.4 |
| B | 26.6 | 9.0 | 0.01 | 3 | 4.4 |

*2.7. Input data processing*

Photos taken with the ARAMIS system camera had the resolution of 2448x2050 pixels. In the case of 256 shades of gray encoded with one byte for each image pixel, this gave a total volume of a single image frame of 5,018,400 bytes. The analysis of 500 photos taken during other study lasted over 10 hours (four threads parallel computation on Core i7 4790K @4.4GHz). However, limiting the area of recorded photos only to the specimen region significantly shortened the computation time (by over 70%). The extraction of an interesting fragment of an example photo is presented in Figure 11.

Test photos present the specimen placed behind two transparent polycarbonate cylinders (Figure 12(a)). Due to those obstacles in the path of the light beam registered by the camera, image corrections related to the light refraction were necessary. A well-known Snell's Law was used for ray tracing [76]

$$\boldsymbol{t} = \frac{\eta_1}{\eta_2}\boldsymbol{i} + \left(\frac{\eta_1}{\eta_2}\cos\theta_i - \sqrt{1-\sin^2\theta_t}\right)\boldsymbol{n}, \qquad (12)$$

$$\boldsymbol{t} = \frac{\eta_1}{\eta_2}[\boldsymbol{n}\times(-\boldsymbol{n}\times\boldsymbol{i})] - \boldsymbol{n}\sqrt{1-\left(\frac{\eta_1}{\eta_2}\right)^2(\boldsymbol{n}\times\boldsymbol{i})\cdot(\boldsymbol{n}\times\boldsymbol{i})}, \qquad (13)$$

where: $\eta_1$ and $\eta_2$ are refractive indices of two materials (air|polycarbonate or polycarbonate|air), $\boldsymbol{i}$ and $\boldsymbol{t}$ are normalized direction vectors of the incident and transmitted rays, $\boldsymbol{n}$ is a normalized vector, orthogonal to the interface and pointing towards the first material (i.e. material with a refractive index $\eta_1$), $\theta_i$ and $\theta_t$ are angles of incidence and refraction, $\times$ and $\cdot$ are cross and dot products, respectively.

To determine the values of the refractive indices $\eta$ for air the approximation formula available in [77] was used (the arguments of this formula are wavelength, temperature, pressure, relative humidity and $CO_2$ concentration). The value of the refractive index for polycarbonate was taken from [78]. Additional image correction was also applied because of the cylindrical side surface of the specimen (Figure 12(b) and 12(c)).

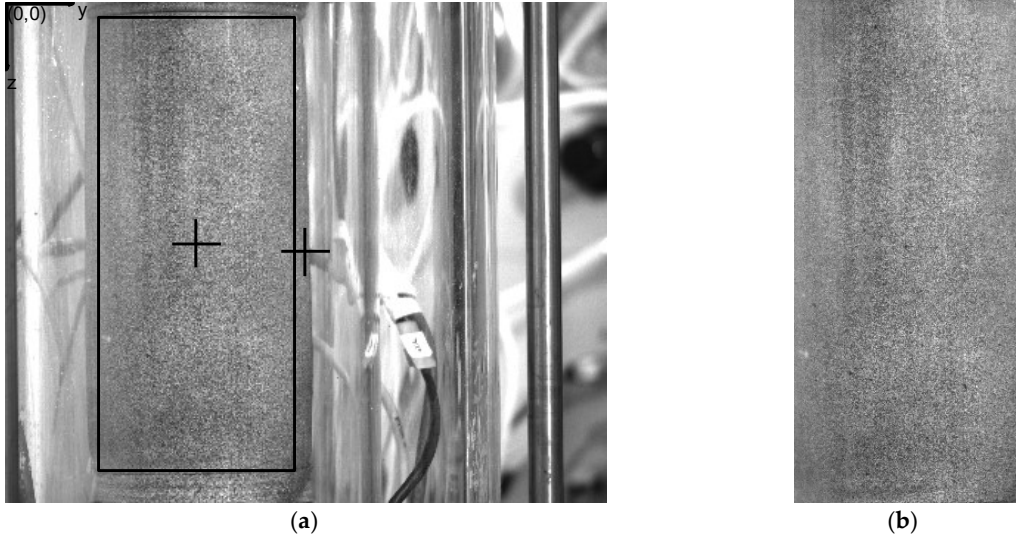

(a)          (b)

**Figure 11.** The area of interest extracted from the photo: (**a**) Original image; (**b**) Extracted area.

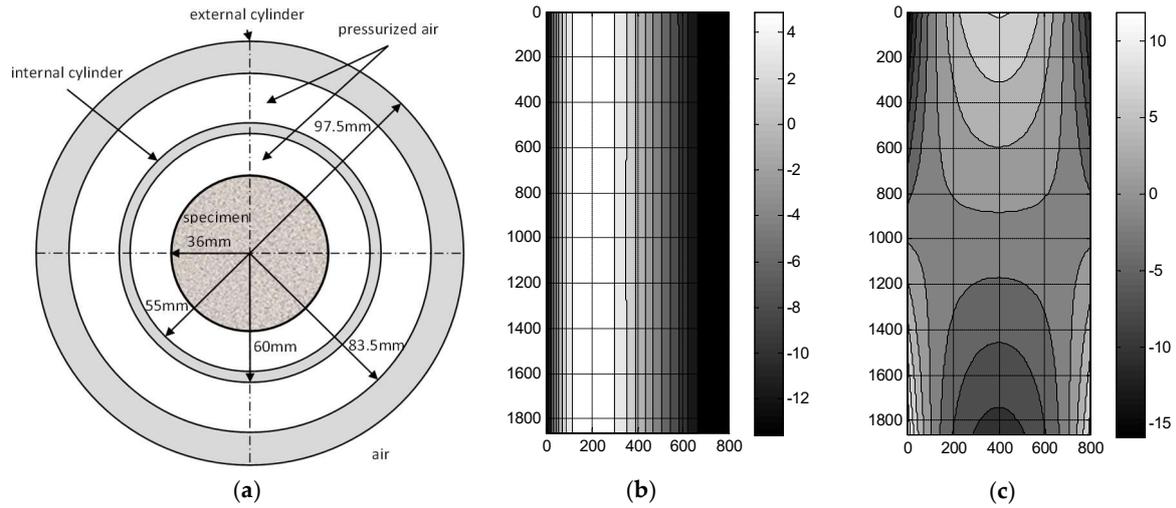

**Figure 12.** (**a**) The cross section of TS device chamber. The cylinders' radii used for calculating refraction correction. (**b**) Width correction map *dy* [mm] for the photo in Figure 11. (c) Height correction map *dz* [mm] for the photo in Figure 11.

Examples of the final correction results are shown in Figures 12(b) and 12(c). It should be emphasized that the application of the above corrections is conditioned by the accuracy of the localization of the camera lens and the RC/TS chamber with the specimen. However, the sensitivity analysis carried out showed that geometric measurements are not critical to the results of the analyzes (i.e. a few centimeter error in measuring the distance of the camera from the specimen's surface does not noticeably affect the results of the deformation analysis).

Another aspect of image analysis is the number of points needed for a comprehensive visualization of deformations. The cropped images contained approximately 1.5 million pixels. Although clear visualization of displacement fields does not require such high resolution, the final analyzes presented in this paper were made for full image resolution.

## 3. Results

The SIFT optical flow code developed in the MATLAB environment by Ce Liu [1] was used in the performed analyzes. The results obtained from optical flow analysis were correlated with known displacements at two reference levels, twisting top cap (measured independently with two proximity sensors) and fixed bottom cap (no displacement). Optical flow algorithm parameters have been optimized as a result of multiple trial and error calculations and are summarized in Table 3 (with their default values suggested by Ce Liu [1]). The algorithm of the full analysis procedure is shown in Figure 13.

**Table 3.** The optical flow algorithm parameters (see [1]).

| Values | Regularization weight | Down-sample ratio | Width of the coarsest level | Number of outer fixed point iterations | Number of inner fixed point iterations | Number of successive over-relaxation iterations |
|---|---|---|---|---|---|---|
| default | 1 | 0.5 | 40 | 3 | 1 | 20 |
| optimal | 0.02 | 0.75 | 20 | 10 | 3 | 40 |

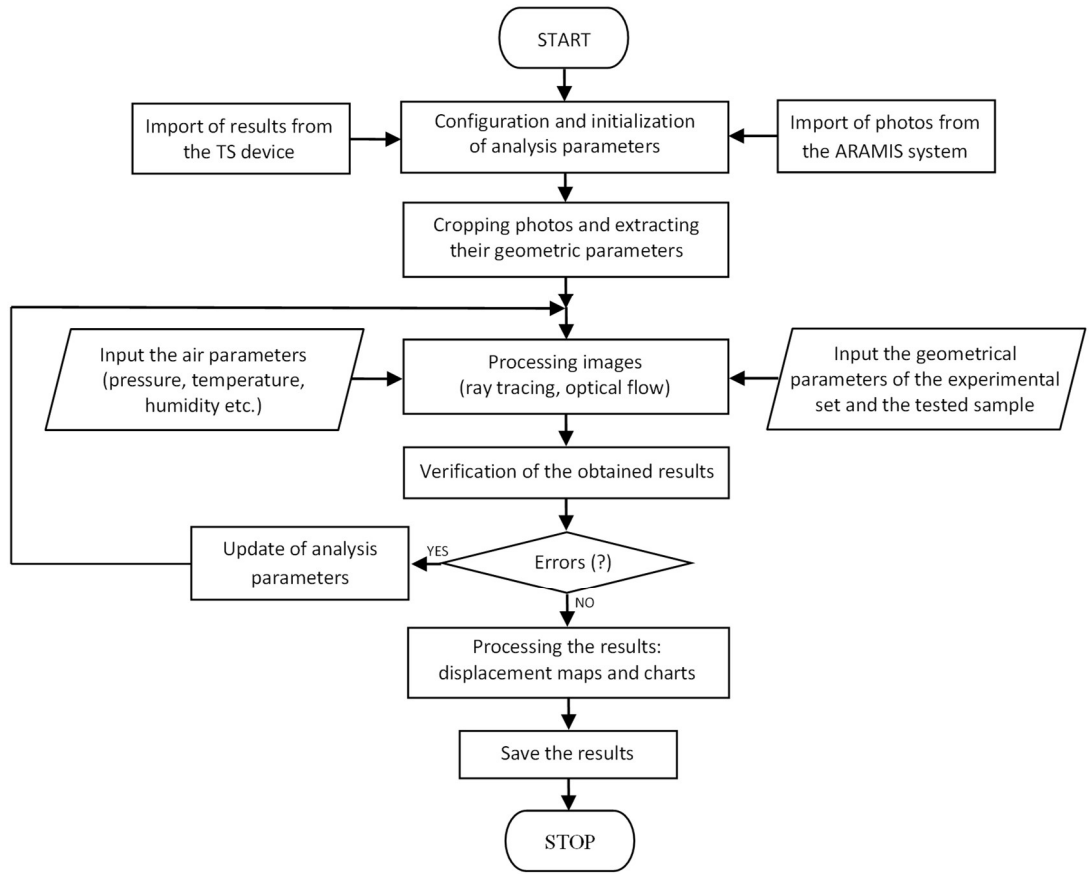

**Figure 13.** The analysis procedure algorithm

The results for the specimen A (subjected to lower confining pressure, see Table 2) are presented on Figures 14-17 in the following manner. Figure 14 shows the torsional shear test results (typical for specimens of this soil tested with similar parameters). Figures 15-17 present the horizontal displacement fields at crucial moments of the TS test, that is the maximum and minimum twist amplitude (first cycle) and the residual state (after 3 cycles and 4.4 h observation time).

The most important goal of the experiments was to determine whether the horizontal displacements of the specimen, observed on its side surface, change linearly with its height. It can be noticed, after averaging the values of horizontal displacements in the band of the assumed width (200 points-pixels) along the specimen axis, that this relationship is not linear (see Figures 15(b), 16(b) and 17(b)Particular attention should be paid to the fact that deviations from the assumed linearity distribute differently on specimen's height during different stages of the same TS test (compare Figures 15(b), 16(b), 17(b)).

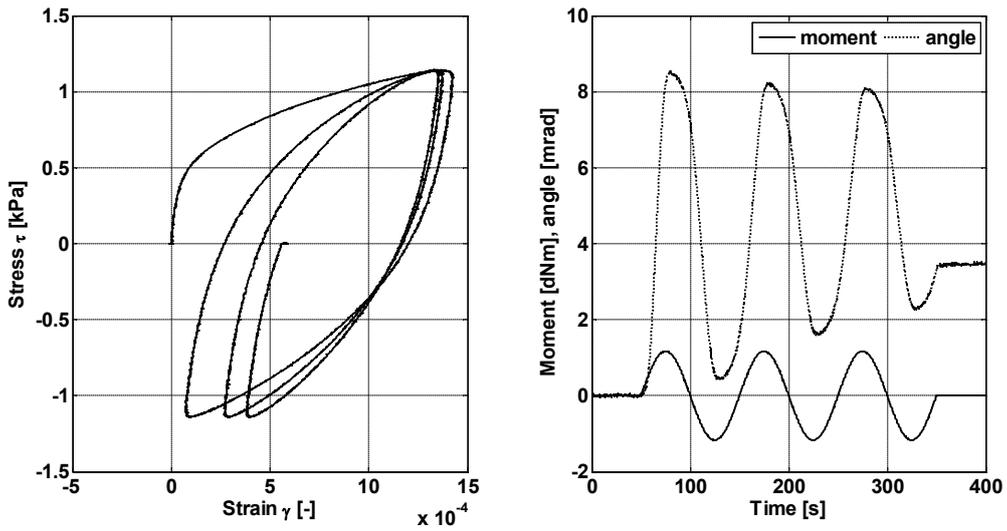

**Figure 14.** Typical TS test results (specimen A)

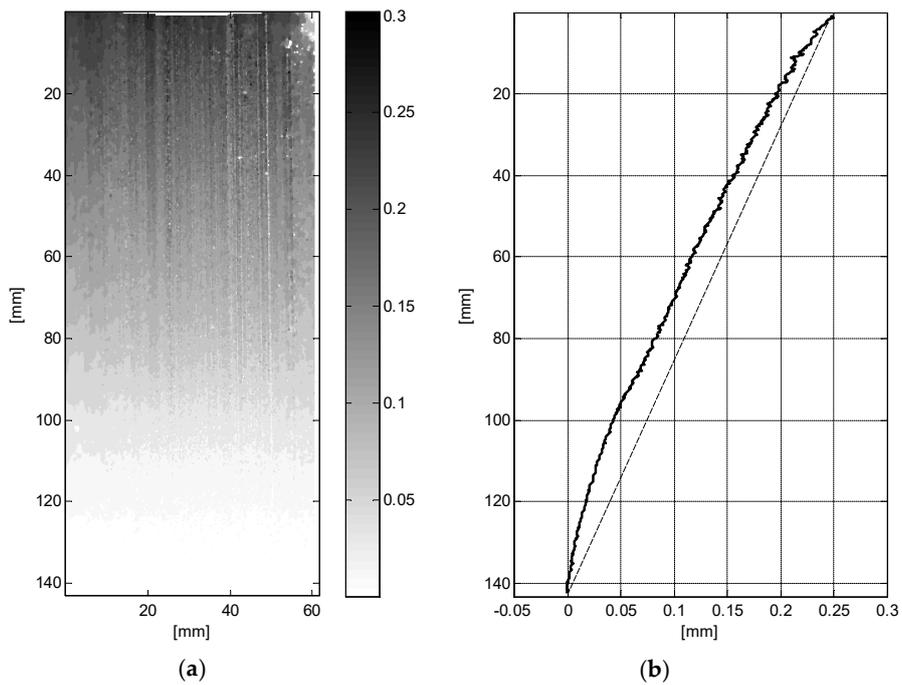

**Figure 15.** Results of optical flow analysis in full resolution (specimen A, grid 801x1861 points). Horizontal displacements at maximum twist (**a**) displacement map (**b**) displacement distribution along the specimen's height (averaging width 200 pixels).

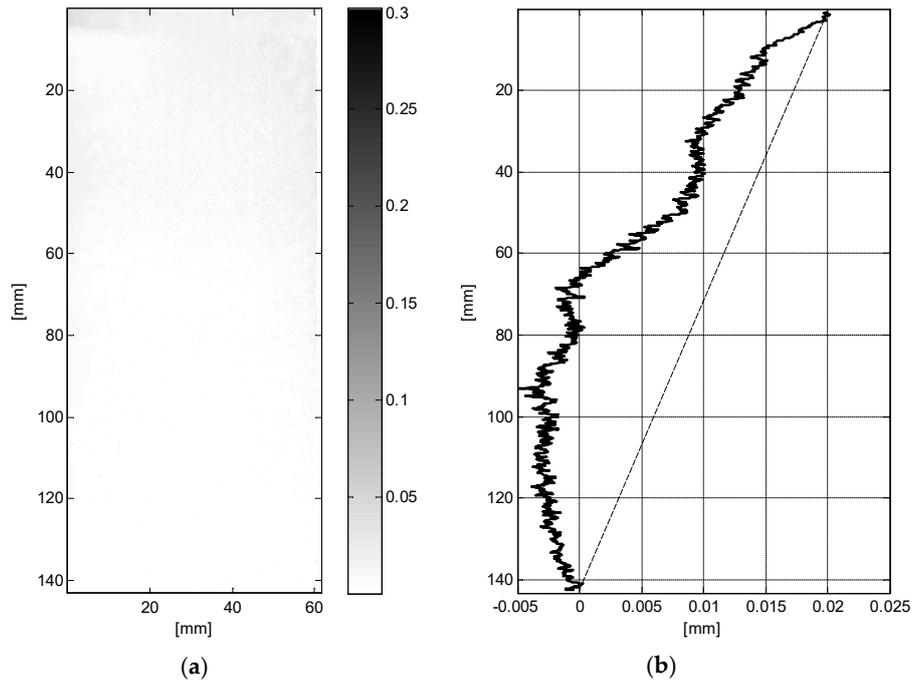

**Figure 16.** Results of optical flow analysis in full resolution (specimen A, grid 801x1861 points). Horizontal displacements at minimum twist (**a**) displacement map (**b**) displacement distribution along the height (averaging width 200 pixels).

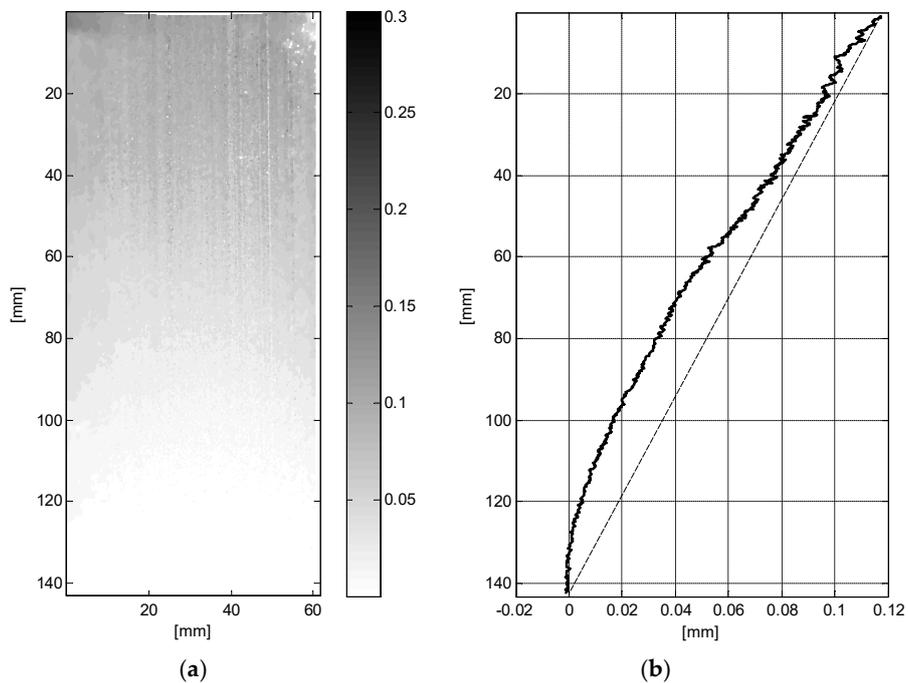

**Figure 17.** Results of optical flow analysis in full resolution (specimen A, grid 801x1861 points). Horizontal displacements in the residual state (**a**) displacement map (**b**) displacement distribution along the height (averaging width 200 pixels).

The results deviate from the assumed linear relationship in terms of shape of the function (clearly non-linear) but they can also lie on either side of that line. An example of this phenomenon can be seen in the results of the TS test carried out on the same sand under a slightly increased isotropic pressure (specimen B, see Table 2), which resulted in an increase in the overall rigidity of the specimen (see Figure 18).

The results for the specimen B are presented in the same manner as for the specimen A. Figures 18-21show the results for the specimen B. Figures 19-21 show a completely different displacement distribution along the height of the specimen (compare to Figures 15-17). It can be noticed that the specimen responds to load differently along its' height and different deformation mechanisms are observed for each value of isotropic pressure.

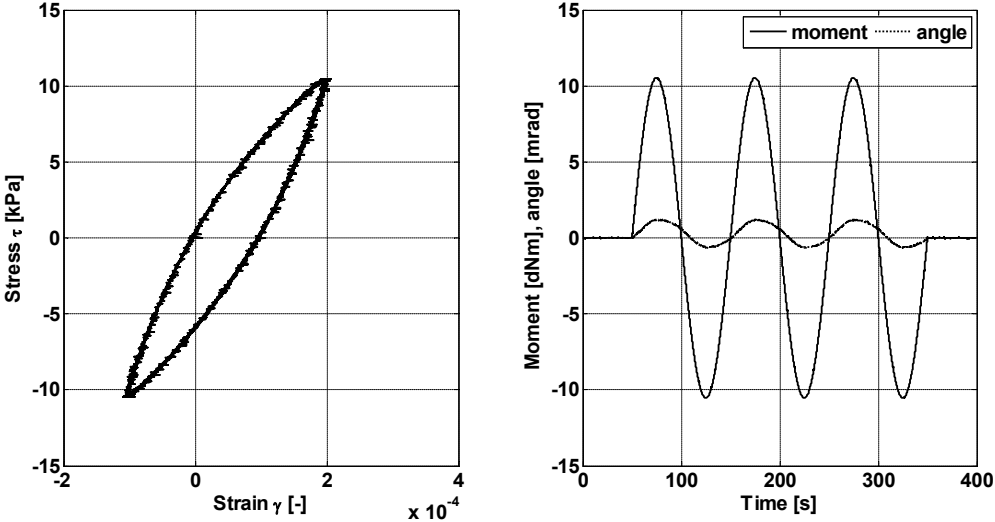

**Figure** 18. Typical TS test results (specimen B)

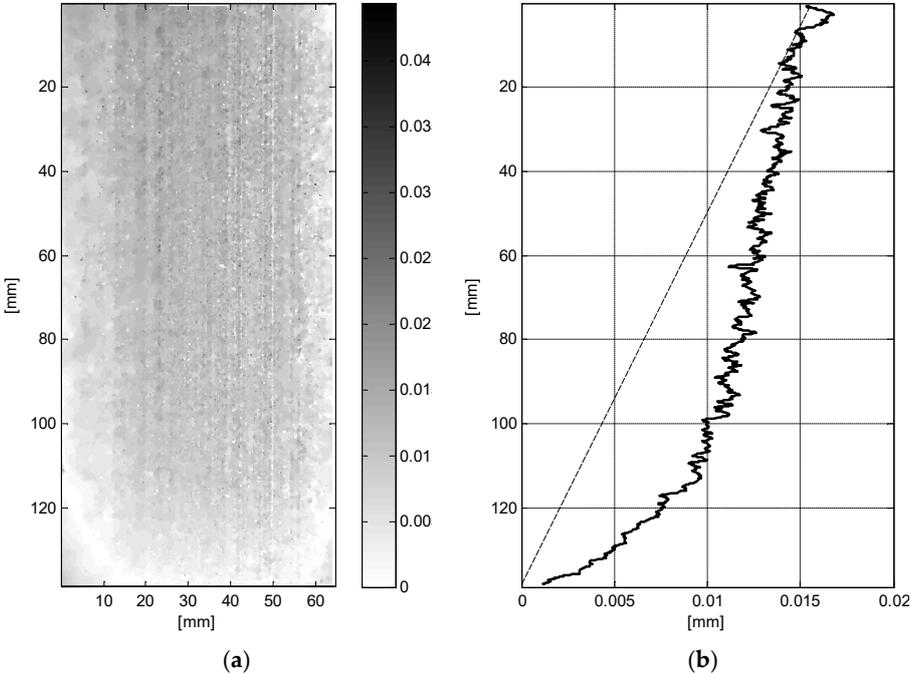

**Figure 19.** Results of optical flow analysis in full resolution (specimen B, grid 801x1861 points). Horizontal displacements at **maximum** twist (**a**) displacement map (**b**) displacement distribution along the height (averaging width 200 pixels).

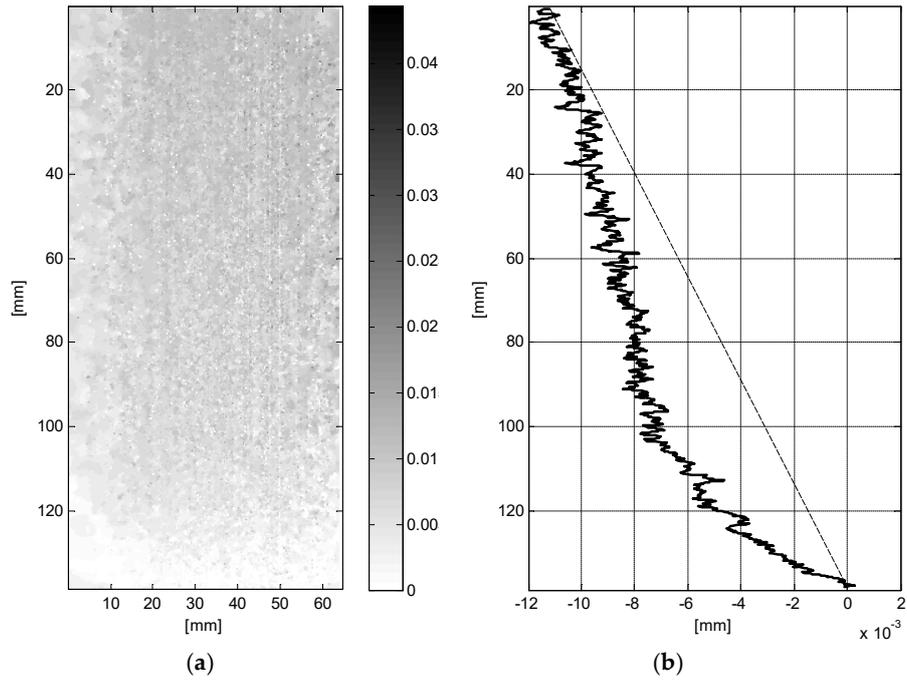

**Figure 20.** Results of optical flow analysis in full resolution (specimen B, grid 801x1861 points). Horizontal displacements at **minimum** twist (**a**) displacement map (**b**) displacement distribution along the height (averaging width 200 pixels).

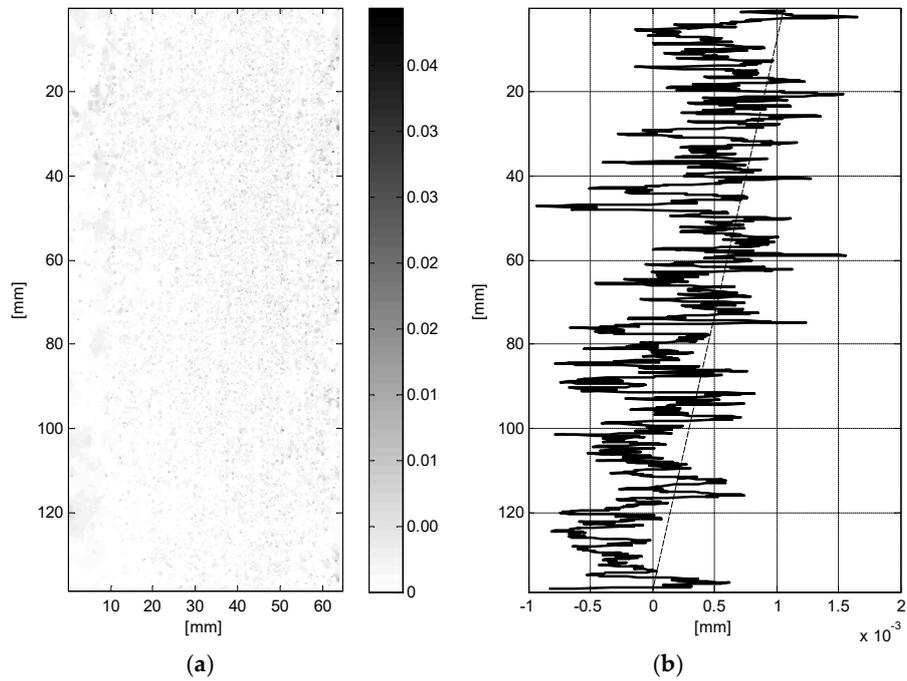

**Figure 21.** Results of optical flow analysis in full resolution (specimen B, grid 801x1861 points). Horizontal displacements in **residual state** (**a**) displacement map (**b**) displacement distribution along the height (averaging width 200 pixels).

The deformation distribution clearly deviates from the assumed linearity for both specimens. Additionally, comparing the results for specimen A and specimen B, two different deformation mechanisms can be observed (see Figure 22).

1. For very low isotropic pressure applied (specimen A) only the top part of the specimen's seem to be activated by torque loading.
2. The higher isotropic pressure applied to specimen B increases specimen's rigidity and larger deformations are observed in the lower part of the specimen.

*3.1. Validation*

Figure 22. shows the results of the horizontal displacement (maximum torque amplitude; Figure 22(a) for specimen A, Figure 22(b) for specimen B) measurement using optical flow method and 3 variants of PIV algorithm. In Table 4 standard deviations are presented for every algorithm. Additionally, Table 5 contains the RMSE for the PIV variants calculated in relation to the optical flow results.

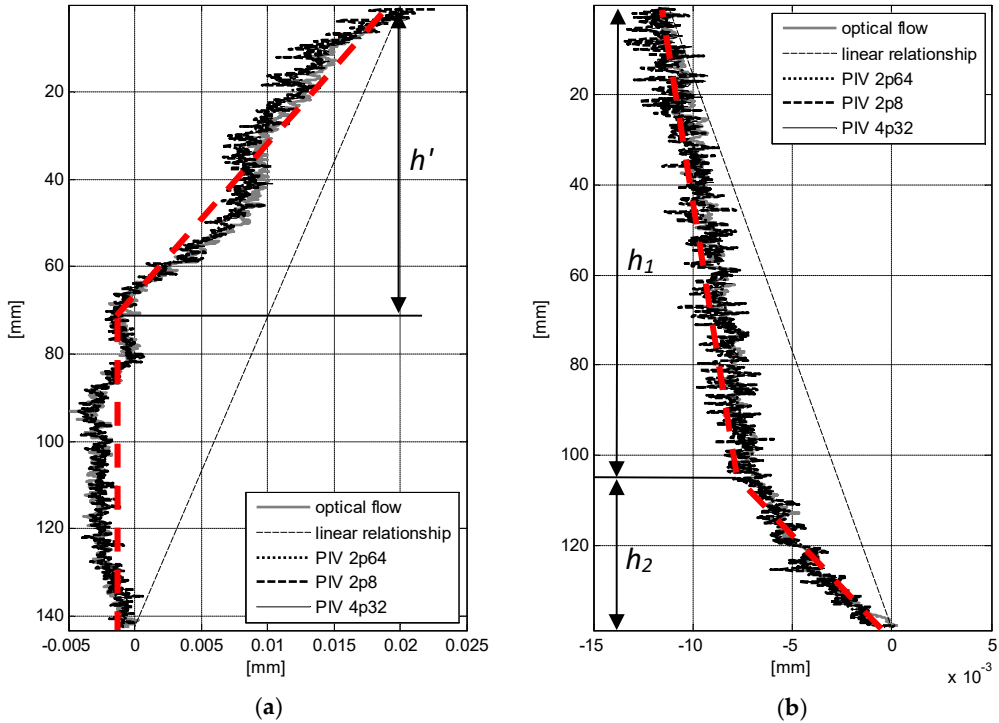

(a) (b)

**Figure 22.** The horizontal displacements distribution along the specimen's height for maximum torque amplitude (a) specimen A, (b) specimen B (full resolution, grid 801x1861 points, averaging width 200 pixels)

Table 4. Standard deviation for the results presented in Figure 22.

| Specimen | Optical flow | PIV 2p64 | PIV 2p8 | PIV 4p32 |
|---|---|---|---|---|
| A | $6.9115*10^{-3}$ | $6.3086*10^{-3}$ | $6.3974*10^{-3}$ | $6.6452*10^{-3}$ |
| B | $2.5682*10^{-3}$ | $2.4346*10^{-3}$ | $2.7313*10^{-3}$ | $2.5238*10^{-3}$ |

**Table 5.** RMSE for the PIV variants calculated in relation to the optical flow results.

| Specimen | PIV 2p64 | PIV 2p8 | PIV 4p32 |
|---|---|---|---|
| A | $7.8589*10^{-4}$ | $10.183*10^{-4}$ | $7.3106*10^{-4}$ |
| B | $3.5910*10^{-4}$ | $7.3712*10^{-4}$ | $4.8697*10^{-4}$ |

It can be seen (Figure 22) that the results obtained with SIFT optical flow and the PIV method show a very good compliance, so the optical flow results can be considered reliable. Similar results were obtained for all of the tested specimens (34) so it can be stated that the observed phenomenon was not occasional.

## 4. Discussion

A direct result of the performed analyzes is a proposed preliminary simplified mechanism of deformation of non-cohesive soil specimens during torsion, presented in Figure 22. This simplification is based on an assumption that the deformation mechanism is bi-linear. In the first case, the modification of assumed deformation distribution is taking into account the shortened height of the specimen $h'$ (Figure 22(a)). The authors propose the term „active height" $h'$. The term refers to the part of the total height $H$ of the soil specimen that undergoes horizontal displacements in reaction to the applied torsional load. The second case (Figure 22(b)) is more difficult to describe due to the strongly non-linear soil stiffness degradation in the range of observed strains. This problem was taken up in further research, the scope of which goes beyond the scope of this work.

It should be particularly stressed, that the observed reduced height that undergoes deformation, the "active height" $h'$, means that the actual shear strain $\gamma$ can be significantly larger than the shear strain commonly calculated with the total specimens' height $H$ (see equation (3)).

The direct result of the above is presented in Figure 23 as hysteresis loops based on the original $\gamma$ value and the $\gamma$ value calculated with $h'$ (modified hysteresis).

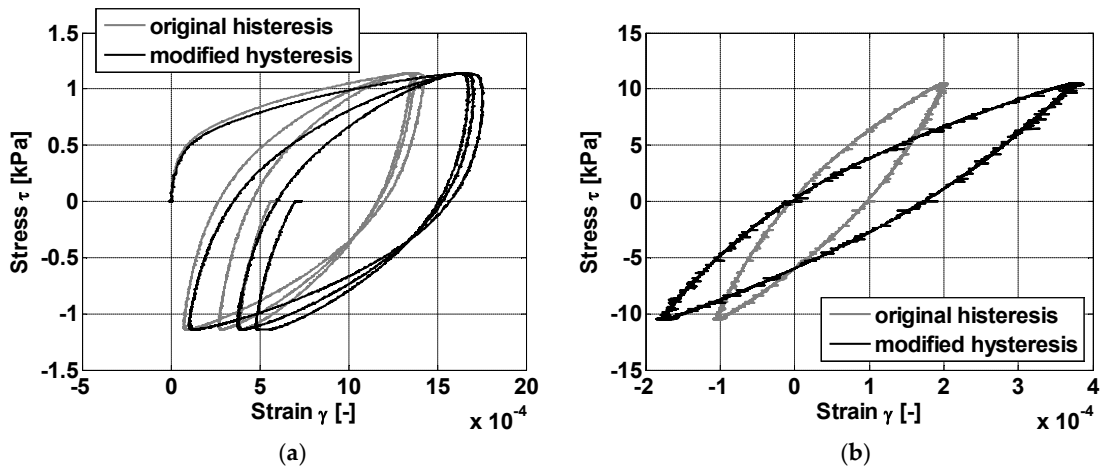

**Figure 23.** The hysteresis loops resulting from the TS device proximity sensors (original hysteresis) and from the optical flow observations (modified hysteresis) (a) specimen A (b) specimen B

What is more, using underestimated strain values means that the determination of shear modulus $G$ value is affected by a fairly high error. As the shear modulus $G$ is one of the most commonly used soil stiffness parameters, the overestimation of the $G$ value can lead to serious calculation errors, e.g. in numerical analyses of soil behavior.

It should be noted that the specimen's vertical displacements also have an impact on the shear modulus $G$ determination as the specimen's rotation angle $\phi$ is divided by specimen's height $H$ to calculate shear strain $\gamma$. However, the vertical displacements obtained from tests were close to 0.01 mm (giving the strains close to 0.001%). The results show that the height of the specimen reacting to applied loading (the "active height" $h'$) is significantly shorter than the specimen's height $H$ even after the vertical displacement occur. Therefore, a conclusion has been drawn that the error emerging form neglecting the influence of vertical displacements on shear modulus determination is minor comparing to the error caused by calculating the shear modulus using the whole specimen's height $H$.

It should also be noted, that performing the torsional shearing soil tests within the small strain range on relatively small sized specimens can potentially result in noticeable local deformations.

When standard TS test result interpretation is applied, the deformation is linearly approximated between fixed bottom and near-the-top proximity sensor readings. This approach can lead to neglecting a possible emergence of local deformation anomalies. Using methods like SIFT optical flow the local phenomena in deformation mechanism can be detected and properly examined.

It is obvious that the optical flow method is sensitive to pixel dimensions in analyzed images. This is well visible in Figures 17 and 21 where the observed displacement to the pixel dimension ratio is large and small, respectively. It should also be noted that the resolution of the results obtained is greater than the dimension of a single pixel in registered images (approx. 0.07 mm), which results from the complex analysis of separated pixel systems and consideration of changes in their brightness. This undoubted advantage can, however, cause a lot of trouble in practice. For instance while recording one particular photo during the tests, someone entered the laboratory and their reflection in the external cylinder of the testing device chamber affected the result image. Additionally, small uncontrolled movement of objects near the system leads to false results. Another problem is the quality of the transparent cylinders covering the RC/TS device. Material flaws invisible to the naked eye are revealed in the results of the analyzes. For example, traces resulting from tailored forming of the inner cylinder can be seen in Figures 15(a) and 19(a) in the form of vertical light and dark lines. Of course, this kind of distortion can be eliminated in the results by using appropriate filters (or eventually manual correction). On the other hand, this sensitivity of the method can be successfully used to identify flaws and imperfections in transparent and reflective materials.

The PIV method used for validation of the results obtained with the SIFT optical flow method. In Figure 22. it can be seen that the results show a very good compliance. However in Figure 24. the advantage of the SIFT optical flow method can be noticed for a series of consequent frames with similar displacements (very small difference in tracked points position). The results obtained with PIV suggest that this method has a limited reliability in case of very small displacement analysis.

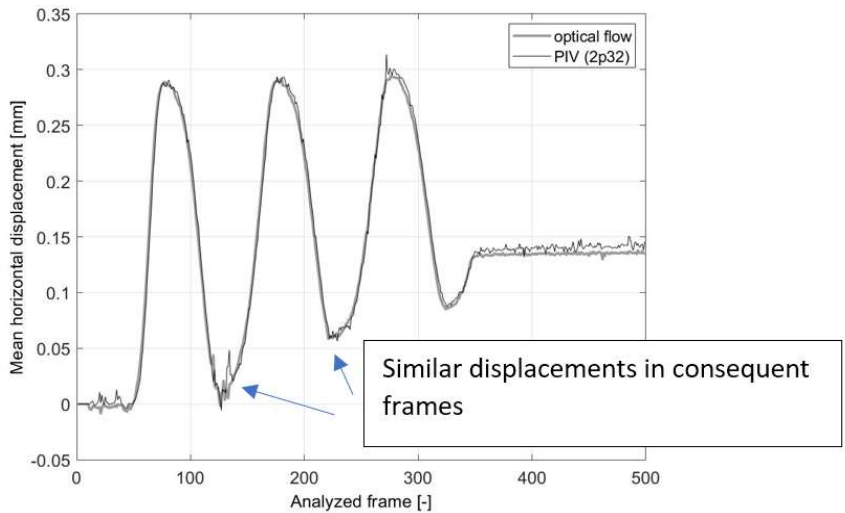

**Figure 24.** Mean horizontal displacement (specimen A) obtained with optical flow and PIV (2p32)

Another issue is the importance of properly selected IA dimensions, which should be adjusted to the expected displacement values. For example, the additional analysis performed with 4p16 configuration (4 passes with 16 pixels, refined to 8, 4 and 2 pixels) resulted in NaN error for all analyzed points.

**Table 6.** Number of analyzed points for variants of PIV method compared to the number of analyzed points for optical flow method

| Specimen | - | 2p64 (0% NaN errors) | 2p8 | 4p32 | 4p16 (100% NaN errors) |
|---|---|---|---|---|---|

| A | Number of points | 5634 | 370670 | 370670 | 1487999 |
| B | Number of points | 5499 | 368045 | 368045 | 1477399 |
| A, B | PIV/optical flow [%] | 0.37 | 24.87 | 24.87 | 99.82 |

Table 6. shows that results not affected by NaN errors are obtained for relatively large IAs. In those cases only a small percentage of points analyzed with SIFT optical flow is tracked. Therefore the PIV method can be problematic when measuring small displacements. Moreover, the small amount of analyzed points makes it impossible to detect potential local anomalies in deformation distribution. It should be noted that for SIFT optical flow those errors were not observed.

The above mentioned issues suggest that the PIV method can be used for small deformation analysis to a limited extend.

The computation time is comparable for both PIV and SIFT optical flow algorithms (3-4 hours for 500 images 800x1800 pixels each).

## 5. Conclusions

The obtained results justify the conclusion that while the displacement values change approximately monotonically along the specimen's height, the shape of displacement function is very different for specimens under different isotropic pressure. Moreover, the deviations from the assumed linearity distribute differently on specimen's height during different stages of the same TS test.

It cannot be unequivocally determined how much of the specimen's height undergoes systematic deformation during TS tests, and therefore authors suggest that measurements of displacements on the side surface of the specimen should always accompany such tests. The correctly determined "active" height $h'$ of the specimen is one of the most important conditions for the correct estimation of the soil stiffness modulus in laboratory tests. The results presented in this paper signal the possibility that the deformation calculated from the rotation angle measured using proximity sensors can be seriously underestimated. As the angle measurements are used for determining the nonlinear stiffness degradation characteristics, e.g. $G(\gamma)$ functions, the error is propagated to the geomechanical modelling and the structural design process.

The authors will continue their work on observing non-linear soil deformations (in small strain range) and taking those observations into account in the torsional shearing tests results interpretation. Specific future research plans are focused on further examination of the observed phenomenon and developing a model which could more accurately describe the occurring mechanisms.

The optical flow techniques are being intensively developed but the progress in measurement of small deformations in soils is constantly insufficient. Therefore, the research team intends to continue searching for optimal measurement conditions and image recording techniques so that the quality of the results obtained from optical flow method is as high as possible and ready to use in engineering practice.

**Acknowledgements**
We would like to express our deep gratitude to the anonymous Reviewers for careful reading of the manuscript and valuable suggestions which greatly improved the presentation of the results obtained in this article.

**Author Contributions:** Conceptualization, P.S. and M.B.[(1)]; methodology, P.S.; software, P.S.; validation, P.S., M.B.[(1)], M.B.[(2)] and R.O.; formal analysis, M.B.[(2)] and R.O.; investigation, M.B.[(1)]; writing—original draft preparation, P.S. and M.B.[(2)]; writing—review and editing, M.B.[(2)]; visualization, M.B.[(1)]; supervision, P.S., M.B.[(2)] and R.O. All authors have read and agreed to the published version of the manuscript.

**Funding:** This research received no external funding.

**Conflicts of Interest:** The authors declare no conflict of interest.